\documentclass[aps,twocolumn,showpacs,preprintnumbers,nofootinbib,floatfix]{revtex4}
\usepackage{graphicx}% Include figure files
\usepackage{epsfig}
\usepackage{dcolumn}% Align table columns on decimal point
\usepackage{subfigure}
\usepackage{CJK}
\usepackage{amsmath}
\def\be{\begin{equation}}
\def\ee{\end{equation}}
\def\bea{\begin{eqnarray}}
\def\eea{\end{eqnarray}}

\def\no{\nonumber}

\newcommand{\omits}[1]{}

\def\bsp{\be\begin{split}}

\def\bes{\be  \begin{split}}

\newcommand{\Rmnum}[1]{\expandafter\@slowromancap\romannumeral #1@}

\def\NPB{{Nucl. Phys. B}}
\def\PRD{{Phys. Rev. D}}

\def\CQG{{Class. Quant. Grav. }}

\begin{document}
\begin{CJK*}{GBK}{song}

\title{Entropy of Reissner-Nordstr\"{o}m-de Sitter black hole}
\author{Li-Chun Zhang$^{a,b}$, Ren Zhao$^{b}$ and Meng-Sen Ma$^{a,b}$\footnote{Email: mengsenma@gmail.com; ms\_ma@sxdtdx.edu.cn}}

\medskip

\affiliation{\footnotesize$^a$Department of Physics, Shanxi Datong
University,  Datong 037009, China\\
\footnotesize$^b$Institute of Theoretical Physics, Shanxi Datong
University, Datong 037009, China}

\begin{abstract}
 Based on the consideration that the black hole horizon and the cosmological horizon of Reissner-Nordstr\"{o}m black hole in de Sitter space are not independent each other, we conjecture the total entropy of the system should have an extra term contributed from the entanglement between the two horizons, except for the sum of the two horizon entropies. Making use of the globally effective first law and the effective thermodynamic quantities, we derive the total entropy and find that it will diverge as the two horizons tends to coincide.

\end{abstract}

 \maketitle

The astronomical observations show that our Universe is probably approaching de Sitter one\cite{AE}. However, as is well known that in de Sitter space there is no spatial infinity and no asymptotic Killing vector which is globally timelike\cite{BBM}.
Moreover, black holes in de Sitter space cannot be in thermodynamic equilibrium in general. Because there are multiple horizons with different temperatures for de Sitter black holes.
To overcome this problem, one can analyze one horizon and take another one as the boundary or separate the two horizons by a thermally opaque membrane or box\cite{AG,Sekiwa}. In this way, the two horizons can be analyzed independently. Besides, one can also take a global view to construct the globally effective temperature and other effective thermodynamic quantities\cite{Urano, Ma}. No matter which method is used, the total entropy of de Sitter black hole is supposed to be the sum of both horizons, namely $S=S_b+S_c$\cite{Kastor:1993}.

We think that the truth may be not so simple because the event horizon and the cosmological horizon are not independent. There may exist some correlations between them due to the following considerations. We can take the Reissner-Nordstr\"{o}m-de Sitter (RNdS) black hole as example.
There are first laws of thermodynamics for both horizons. According to \cite{Dolan}, the first laws for the black hole horizon and the cosmological horizon are respectively:
\bea\label{1st}
dM&=&T_{b}dS_{b}+\Phi_b dQ+V_bd\Lambda,\no \\
dM&=&-T_{c}dS_{c}+\Phi_c dQ+V_cd\Lambda,
\eea
where $M$ is the conserved mass in dS space, $\Phi$ and $Q$ stand for the electric potential and electric charge, and $\Lambda$ is the cosmological constant. Including the variable $\Lambda$, the above first laws can have corresponding Smarr formulae, which are also given in \cite{Dolan}.
The two laws in Eq.(\ref{1st}) are not truly independent. They depend on the same quantities $M,~Q,~\Lambda$.
All the geometric and thermodynamic quantities for the both horizons can be represented by $M,~Q,~\Lambda$. Therefore, the size of black hole horizon is closely related to the size of the cosmological horizon, and the evolution of black hole horizon will lead to the evolution of the cosmological horizon.

Considering the correlation or entanglement between the event horizon and the cosmological horizon, the total entropy of the RNdS black hole is no longer simply $S=S_b+S_c$, but should include an extra term from the contribution of the correlations of the two horizons.

The line element of the RNdS black holes is given by
\be\label{staticmetric}
ds^2=-h(r)dt^2+h(r)^{-1}dr^2 + r^2d\Omega^2,
\ee
where
\be\label{fr}
h(r) = 1 - \frac{{2M}}{r} + \frac{{{Q^2}}}{{{r^2}}} - \frac{\Lambda }{3}{r^2}.
\ee
There are three positive real roots for $h(r)=0$. The smallest one $r_{-}$ is the inner/Cauchy horizon,
the intermediate one $r_{+}$ is the event horizon of black hole and the largest one is the cosmological horizon.

Generally, the temperatures at the event horizon and cosmological horizon are not equal. Thus, the whole RNdS system cannot be in equilibrium thermodynamically.
However, there are two special cases for the RNdS black hole, in which the temperatures at the both horizons are the same. One case is the so-called Nariai black hole, the other is the lukewarm black hole\cite{Romans,Kastor,Cai:1998,Huang}. For the Nariai black hole, the event horizon and cosmological horizon coincide \emph{apparently} and have the same temperature, zero or nonzero\cite{Hawking}.

Considering the connection between the black hole horizon and the cosmological horizon, we can derive the effective thermodynamic quantities and corresponding first law of black hole thermodynamics\footnote{One can also
take the cosmological constant $\Lambda$ as the pressure and then derive the effective volume. In this case the effective first law should be: $dM={T_{eff}}dS  + {\phi _{eff}}dQ + {V_{eff}}dP$. This has been done in another  paper.}:
\be\label{eff1st}
dM = {T_{eff}}dS - {P_{eff}}dV + {\phi _{eff}}dQ.
\ee
Here the thermodynamic volume is that between the black hole horizon and the cosmological horizon, namely
\be\label{Vol}
V = \frac{{4\pi }}{3}\left( {r_c^3 - r_ + ^3} \right) = \frac{{4\pi }}{3}r_c^3\left( {1 - {x^3}} \right),
\ee
where $x=r_{+}/r_{c}$.

The total entropy can be written as
\be\label{ent}
S = {S_ + } + {S_c} +S_{ex}= \pi r_c^2\left[1 + {x^2} + f(x)\right].
\ee
Here the undefined function $f(x)$ represents the extra contribution from the correlations of the two horizons.
Then, how to determine the function $f(x)$?

In general cases, the temperatures of the black hole horizon and the cosmological horizon are not the same, thus the globally effective temperature $T_{eff}$ cannot be compared with them. However, in the special case, such as lukewarm case, the temperatures of the two horizons are the same. We conjecture that in this special case the effective temperature should also take the same value. On the basis of this consideration, we can obtain the information of $f(x)$.

According to Eq.(\ref{eff1st}), the effective temperature can be derived by
\bea
{T_{eff}} &=& {\left( {{{\partial M} \over {\partial S}}} \right)_{Q,V}}  \\
&=&{{{1 \over {{r_c}}}{{\left( {{{\partial M} \over {\partial x}}} \right)}_{{r_c}}}\left( {1 - {x^3}} \right) + {{\left( {{{\partial M} \over {\partial {r_c}}}} \right)}_x}{x^2}} \over {2\pi {r_c}\left[ {(2x + f'(x))(1 - {x^3})/2 + (1 + {x^2} + f(x)){x^2}} \right]}}. \no
\eea
From Eq.(\ref{fr}), one know that
\be\label{mass}
M=\frac{(x+1) \left(x^2 r_c^2+Q^2 x^2+Q^2\right)}{2 x \left(x^2+x+1\right) r_c}.
\ee
We can derive the effective temperature
\bea
T_{eff}&=&\frac{1}{2\pi x^2 A(x)}\left[ { Q^2 \left(x^6+x^5+x^4-2 x^3+x^2+x+1\right)  }\right. \no \\
      &-& \left.{x^2 \left(x^4+x^3-2 x^2+x+1\right) r_c^2 }\right],
\eea
where
\bea
A(x)&=&r_c^3 \left(x^2+x+1\right) \left[\left(x^3-1\right) f'(x)-2 x^2 f(x) \right.\no \\
    &-& \left. 2 (x+1) x\right].
\eea

In the lukewarm case, there is
\be
M^2=Q^2=\frac{x^2 r^2_c}{(x+1)^2}.
\ee
With this, we can obtain the effective temperature of lukewarm RNdS black hole:
\be
T_{eff}=-\frac{x \left(x^4+1\right)}{\pi  (x+1)^2 \left(x^2+x+1\right) r_c A(x)}.
\ee

We also know that for the lukewarm RNdS black hole, the temperature is
\be
T_{+}=T_c=\frac{1-x}{2 \pi  \left(1+x\right)^2 r_c}.
\ee
Equating the two temperatures, we obtain a differential equation for $f(x)$:
\bea
f'(x)+\frac{2 x^2 }{1-x^3}f(x)=\frac{2x^2 \left(2 x^3+x^2-1\right)}{\left(1-x^3\right)^2}.
\eea
Fortunately, this equation have analytic solution, which is

\bea
f(x)=\frac{8}{5} \left(1-x^3\right)^{2/3}-\frac{2 \left(x^5+5 x^3-4\right)}{5 \left(x^3-1\right)},
\eea
where we have taken the boundary condition $f(0)=0$. 

\begin{figure}[!htbp]
\center{ \subfigure[]{ \label{1-a}
\includegraphics[width=7cm,keepaspectratio]{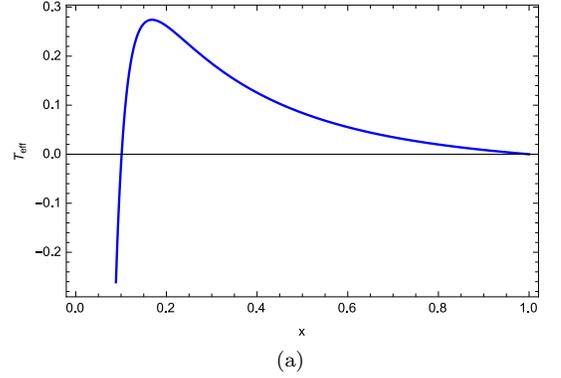}\hspace{0.5cm}}\\
\subfigure[]{ \label{1-b}
\includegraphics[width=7cm,keepaspectratio]{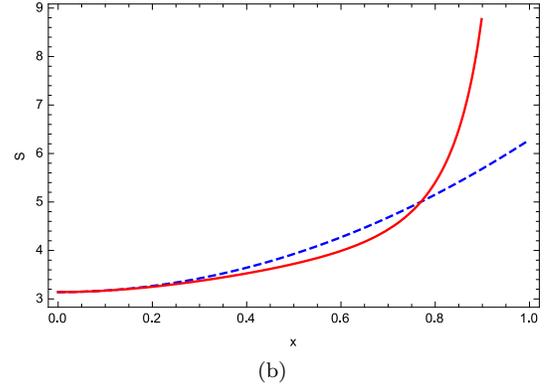}}
\caption{$T_{eff}$ and total entropy $S$ with respect to $x$. In (a), $T_{eff}$ has a maximum at $x=0.181$. In (b), the dashed (blue) curve represents the sum of
 the two horizon entropy and  the solid (red) curve depicts the result in Eq.(\ref{ent}). we set $r_c=1$ and $Q=0.1$.\label{fx}}
}
\end{figure}

In Fig.\ref{fx}, we depict the effective temperature $T_{eff}$ and $S$ as functions of $x$. It is shown that $T_{eff}$ tends to zero as $x \rightarrow 1$, namely the charged Nariai limit.
Although this result does not agree with that of Bousso and Hawking\cite{Hawking}, it is consistent with the entropy. As is depicted in Fig.\ref{1-b}, the entropy will diverge as $x \rightarrow 1$.
Besides, one can see the entropy is monotonically increasing with the increase of $x$, while $T_{eff}$ first increases and then decreases. According to the general definition of heat capacity, $C=\frac{\partial M}{\partial T}=T\frac{\partial S}{\partial T}$, only in the region of $x$ with positive temperature and positive slope the RNdS black hole can be thermodynamically stable. This is unexpectable. This means when the black hole horizon and the cosmological horizon are too far away (small $x$) or too close (large $x$), RNdS black hole cannot be thermodynamically stable.

In this letter, we have presented the entropy of RNdS black hole. It is not only the sum of the entropies of black hole horizon and the cosmological horizon, but also with an extra term from the correlation between the two horizons. This idea has twofold advantages. First, if without the extra term in the total entropy, the effective temperature is not the same as that of the black hole horizon and the cosmological horizon in the lukewarm case. This is not satisfactory. Second, the method of effective first law of thermodynamics lacks physical explanation or motivation. While taking advantage of the method, we obtain the corrected entropy of RNdS black hole, which may make the method more acceptable.

\section*{Acknowledgements}
This work is supported in part by the National Natural Science Foundation of China under Grants
 Nos.(11475108) and by the Doctoral Sustentation Fund of Shanxi Datong
University (2011-B-03).

\end{CJK*}

\end{document}